\documentclass[runningheads]{svmult}

\usepackage{makeidx}   
\usepackage{graphicx}  
\usepackage{subeqnar}  
\usepackage{multicol, epsfig}  
\usepackage{physprbb}  
\makeindex             

\newcommand{\gam}{$\gamma$}
\newcommand{\about}{$\sim$}

\begin{document}
\title*{The GLAST Burst Monitor (GBM)}
\toctitle{A Gamma-Ray Burst Monitor for GLAST}
%
%
\titlerunning{A Gamma-Ray Burst Monitor for GLAST}
%
\author{G. G. Lichti\inst{1}
\and M. S. Briggs\inst{3}
\and R. Diehl\inst{1}
\and G. Fishman\inst{2}
\and R. Georgii\inst{1}
\and R. M. Kippen\inst{3}
\and C. Kouveliotou\inst{2}
\and C. Meegan\inst{2}
\and W. Paciesas\inst{3}
\and R. Preece\inst{3}
\and V. Sch\"onfelder\inst{1}
\and A. von Kienlin\inst{1}}
\authorrunning{G. Lichti et al.}
%
%
\institute{Max-Planck-Institut f\"ur extraterrestrische Physik, 85748 Garching
\and NASA/Marshall Space-Flight Center, Mail Code SD50, Huntsville, AL 35812
\and University of Alabama in Huntsville, AL 35899}

\maketitle              

\begin{abstract}
The selection of the GLAST burst monitor (GBM) by NASA will allow the
investigation of the relation between the keV and the MeV-GeV emission from
$\gamma$-ray bursts. The GBM consists of 12 NaI and 2 BGO crystals allowing a
continuous measurement of the energy spectra of $\gamma$-ray bursts from
$\sim5$ keV to $\sim30$ MeV. One feature of the GBM is its high time
resolution for time-resolved $\gamma$-ray spectroscopy. Moreover the
arrangement of the NaI-crystals allows a rapid on-board location
($<15^\circ$) of a $\gamma$-ray burst within a FoV of $\sim8.6$ sr. This position
will be communicated to the main instrument of GLAST making follow-up
observations at high energies possible.
\end{abstract}

\section{Introduction}

It was in 1994 that EGRET observed a \gam-ray burst which showed a \gam-ray
emission above 50 MeV up to \about 1.5 hours after the start of the burst.
The \gam-quantum with the highest energy was observed after \about 1.3 hours
with an energy of 18 GeV \cite{Hurley94}. This was an unexpected and very
surprising result and as of yet the relation between this high-energy
and the low-energy emission is not understood. It is a goal of the
GLAST mission to investigate this relation and to unravel the underlying physical processes.

\section{Characteristics of energy spectra of \gam-ray bursts}

A typical spectrum of a \gam-ray burst is characterized by a broken power law
(see Figure 1). Below a certain break energy $E_{\rm p}$ the spectrum can be
described by a power law E$^{\alpha}$ with an exponential decline, whereas
above this energy it is a pure power law E$^{\beta}$ indicating a non-thermal
origin of this part of the spectrum. The break-energy $E_{\rm p}$ at which the
luminosity reaches a maximum has a log-normal distribution around an energy of
250 keV \cite{Preece00}.
In order to determine $E_{\rm p}$ well enough one needs a long lever arm
on both sides of this energy, accentuating the importance of low- and
high-energy measurements.

\begin{figure}
\begin{center}
\psfig{figure=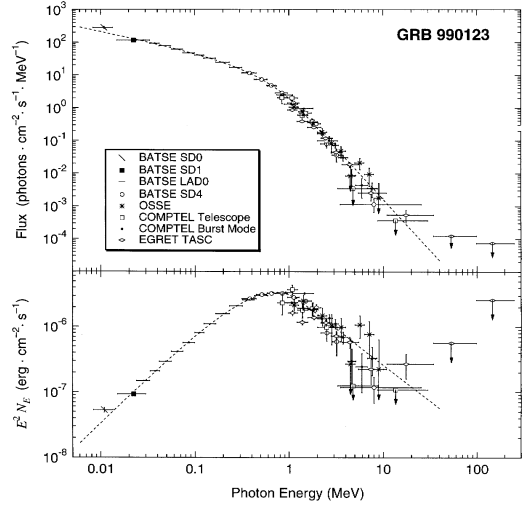,width=7cm,clip=}
\end{center}
\caption{The energy spectrum of GRB990123}
\label{labelfig1}
\end{figure}

The spectral index $\alpha$ of the low-energy emission is distributed around
a value of -1 spanning the range from -2 to 0. This distribution strongly
constrains popular burst-emission models like the synchrotron emission from
shocked electrons (\cite{Tavani96} and \cite{Rees92}) or the blast-wave model
\cite{Cen99} strongly. The distribution of the spectral index $\beta$ of the
high-energy emission reaches a maximum around -2.3. It extends from -3.5 to
-1.4 \cite{Preece00}.
The spectra with $\beta$-values $>$-2 are of special interest because
in this case the spectrum would diverge for E $\rightarrow$ $\infty$. Therefore
a cut off at high energies must exist. An interersting question which will be
answered by GLAST is, at which energy this cutoff occurs.

\section{The GLAST mission}

GLAST will continue the successful measurements of EGRET in a wider
energy range, with a higher sensitivity and with a better location accuracy.
Its main instrument, the Large-Area Telescope (LAT), will use the same
physical processes as EGRET for the detection of \gam-rays, but using an
advanced  technical concept. It will consist of an array of towers of
pair-conversion chamber stacks made from Silicon-strip detectors.

The LAT measures \gam-rays in the energy range \about15 MeV to \about500 GeV,
reaching a point-source sensitivity of better than $4 \cdot 10^{-9}$
photons/(cm$^2$ s) above 100 MeV for an observation time of one year.
It will therefore be more than 30 times more sensitive than EGRET. Within its
large FoV of $<$3 sr it will locate point sources from 5' down to 30".
With its fairly good energy resolution of \about 10\% it will measure the
energy spectra of sources with a high accuracy. The LAT is devoted to the
study of particle acceleration in the universe as it takes place in the nuclei
of active galaxies, at or near pulsars, in supernova remnants and in
interactions of the cosmic rays with the interstellar matter. In addition,
it will detect $\sim$50--150 \gam-ray bursts per year. For a description of
GLAST see \cite{Gehrels99}.

From the latter measurements it is hoped to solve the afore-mentioned problem
of the high-energy burst emission. However, the LAT suffers
from some deficiencies because high-energy measurements alone do not allow a
unique classification of \gam-ray bursts, because the break energy $E_{\rm p}$ of a
burst lies below GLAST's energy threshold of \about15 MeV. Therefore without
low-energy measurements a classification of the bursts is difficult, since
most bursts were measured by BATSE at these energies and the connection to the
BATSE data archive cannot be established. Another deficiency is that the
trigger conditions of the LAT for weak bursts are unfavourable because of the
rather high background rate. In order to overcome these deficiencies a
secondary instrument was proposed, the GBM. It will
extend the energy range of GLAST towards lower energies and it will have a
much larger FoV than the LAT. Therefore it will detect more bursts than this
instrument. The GBM will communicate the positions of these bursts to the LAT
which then will, after reorientation if needed, search for
high-energy \gam-rays. Moreover, for weak bursts the LAT will use GBM-provided
information to reduce background by eliminating events with directions far
frm the GBM burst location.

\section{Description of the GBM and its performance}

The goals of the GBM described above can be achieved by an arrangement of 12
thin NaI discs which are oriented such that from the relative count
rates the direction to the burst can be derived (KONUS/BATSE principle). They
will in addition measure the burst spectra in the energy range 5 keV - 1 MeV.

In order to obtain spectral overlap with the LAT, two cylindrical BGO crystals will be
mounted to the GLAST spacecraft which are sensitive to \gam-rays in the
energy range from 150 keV - 30 MeV. A more detailed description of the GBM
can be found in \cite{vKienlin00}. Within the large FoV of \about8.6 sr
\about 215 bursts/year will be detected by the GBM. Most of them will be
located on board in real time with an accuracy $<15^{\circ}$. On ground much
better locations (\about$3^{\circ}$) can be derived. The 50--300 keV sensitivity
for nominal on-board triggers will be \about0.6 photons/(cm$^2$ s), whereas
an ultimate 5$\sigma$ sensitivity of \about0.35 photons/(cm$^2$ s) can be
achieved on ground.

\section{Scientific goals and expected results}

With the GBM continuous measurements of energy spectra from \about 5 keV to
\about 30 MeV will be performed. Apart from spectra also the light curves of
bursts will be recorded with a time resolution in the $\mu$s-range. The GBM will
serve as a sensitive burst trigger for the LAT and will communicate very
rapidly ($<$5 s) the burst location to it. This trigger will initialize
data-reduction modes in the LAT which will then observe the burst and localize
it with a much better accuracy ($<$3'). This precise location will be communicated
within less than 10 s to ground in order to allow a search for objects at
other wavelengths. The burst data collected by the GBM will preserve the
continuity to the BATSE data and enlarge this important archive. Moreover the
GBM will be part of the IPN as a near-earth burst detector.

\begin{figure}
\begin{center}
\psfig{figure=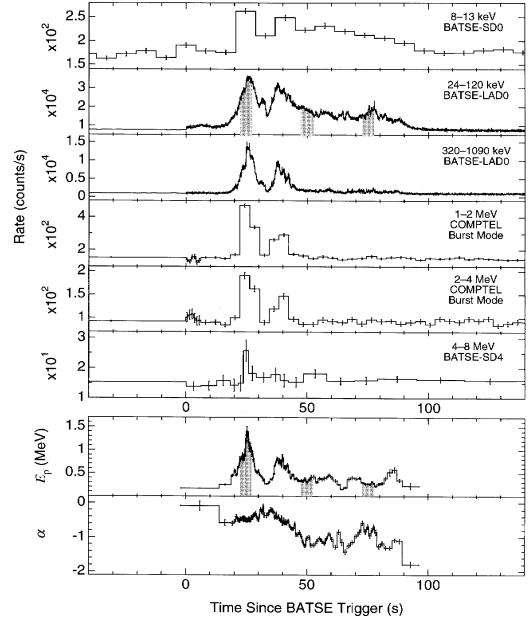,width=7cm,clip=}
\end{center}
\caption{The energy-resolved light curves of GRB990123 and the evolution of
$E_{\rm p}$ and $\alpha$}
\label{labelfig2}
\end{figure}

With the GBM the relation between the keV-MeV-GeV emission can be investigated
in great detail. Time-resolved energy spectra will be measured allowing
time-resolved spectroscopy (see Figure 2 of \cite{Briggs99}). This permits the
investigation of the hard-to-soft evolution of the power-law index $\alpha$
and the hardness-intensity correlation and tackles the problem of the
narrowing of the peaks with energy. It may also give an answer to the question
why the low-energy emission lasts longer than the high-energy one. Together
with the LAT it will be possible to investigate these correlations to high
energies with the aim to measure the suspected cutoff and to find a possible
evolution of the spectral index $\beta$.


\begin{thebibliography}{8.}
\addcontentsline{toc}{section}{References}

\bibitem{Briggs99} Briggs, M. S. et al., Ap. J. \textbf{524}, 82, 1999
\bibitem{Cen99} Cen, R., Ap. J. L. \textbf{517}, L113, 1999
\bibitem{Gehrels99} Gehrels, N. et al., Astroparticle Physics \textbf{11}, 277,
                    1999
\bibitem{Hurley94} Hurley, K. et al., Nature \textbf{372}, 652 (1994)
\bibitem{Preece00} Preece, R. D. et al., Ap. J. Suppl. Ser. \textbf{126}, 19-36,
                   2000
\bibitem{Rees92} Rees, M. J. and P. Meszaros, MNRAS \textbf{258}, 41, 1992
\bibitem{Tavani96} Tavani, M., Ap. J. \textbf{466}, 768, 1996
\bibitem{vKienlin00} von Kienlin, A. et al., Proc. 4$^{th}$ INTEGRAL Workshop,
         Alicante, 2000

\end{thebibliography}
\end{document}